\documentclass{aa}
\usepackage{graphics}
\usepackage{amssymb}

\begin{document}

\title{RR Lyrae stars in the Sagittarius dwarf galaxy: Period analysis\thanks{Based on observations obtained at the European Southern Observatory, La Silla, Chile}}

\author{P. Cseresnjes \inst{1,2}}

\offprints{P. Cseresnjes, \email{patrick.cseresnjes@obspm.fr}}

\institute{DASGAL, Observatoire de Paris, 61 Avenue de l'Observatoire, F-75014 Paris
      \and Centre d'Analyse des Images - INSU}

\date{Received 19 April 2001/ Accepted 1 June 2001}

\abstract{We carried out a period analysis on $\sim$3\,700 RR Lyrae stars on two Schmidt fields centred on (l,b)=(3.1$^{\circ}$,$-$7.1$^{\circ}$) 
and (6.6$^{\circ}$,$-$10.8$^{\circ}$) respectively, covering an area of $\sim$50 deg$^{2}$. These stars are distributed almost evenly between 
the Sagittarius dwarf galaxy (Sgr) and the Milky Way. For Sgr members, the average periods are $\langle P_{ab}\rangle=0.574^{\rm d}$ and 
$\langle P_{c}\rangle=0.322^{\rm d}$ for RRab and RRc stars respectively. This places Sgr in the long-period tail of the 
Oosterhoff I group. We report the detection of 53 double-mode RR Lyrae stars (RRd) within 
our sample. The magnitude of 40 of these stars 
is consistent with membership in Sgr whereas 13 RRds are located within our Galaxy. 
We also found 13 RR Lyraes (5 in Sgr and 8 in the Galaxy) exhibiting two closely spaced frequencies, most probably related to 
non-radial pulsations.
The period distribution of the RR Lyrae 
variables in Sgr is compared to those of other Milky Way satellites. We find a remarkable similarity between the RR Lyrae 
populations in Sgr and the Large Magellanic Cloud (LMC), suggesting that these galaxies have similar horizontal branch morphologies. 
This may indicate that Sgr 
and the LMC started their formation under similar conditions.
Using various photometric indicators, we estimate the metallicity of the RR Lyrae stars in Sgr and find 
$\langle {\rm [Fe/H]}\rangle \simeq-1.6$ 
dex with a dispersion of $\sim \pm$0.5 dex around this value and a minor but significant population at $\lesssim$$-$2.0dex. 
We do not find evidence for a spatial metallicity 
gradient in the RR Lyrae population of Sgr. From the spatial distribution of RR Lyraes, we find that the main body of Sgr contains 
$\sim$4200 RRab stars. Assuming that population gradients are negligible in Sgr, we find M$_{\rm V}$(Sgr)$\simeq -$13.9$^{+0.4}_{-0.6}$ 
mag for the main body. If Sgr has been stripped of 50$\%$ of its mass through Galactic tides, as assumed by some models, it would imply 
a total absolute magnitude of $\sim-$14.7mag for this galaxy. Such a luminosity would be consistent with the empirical 
metallicity/luminosity relation for dwarf spheroidal galaxies.
\keywords{Stars: horizontal-branch - Stars: population II - Stars: variables: RR Lyr - 
Galaxies: dwarf - Galaxies: individual: Sagittarius dwarf - Local group}}
\maketitle
%
%
\section{Introduction}
\indent 
Since RR Lyrae stars are low mass helium core-burning stars, they
 are believed to be older than 10 Gyr. The study of the period distributions of these variables can therefore provide insights into the 
original populations of their host system. 
Field and cluster RR Lyraes show a clear separation between long period/metal poor and short period/metal rich 
stars, known as the Oosterhoff dichotomy (Oosterhoff \cite{oosterhoff}). On the other hand, the average period of RR Lyrae 
populations in Dsph galaxies is intermediate between the two Oosterhoff groups. 
It has been shown that this intermediate status is not the consequence of a superposition of two populations 
 but rather an intrinsic property of RR Lyrae populations in these systems (e.g. Sextans: Mateo et al. \cite{mfk}, 
Leo II: Siegel \& Majewski \cite{sm}). \\
\indent Of particular interest among RR Lyrae stars are those pulsating simultaneously in the 
fundamental and first overtone mode (RRd). 
These stars offer the opportunity to constrain their mass and luminosity independent of stellar evolution theory 
(e.g. Bono et al. \cite{bccm}, Kov\'acs \& Walker \cite{kw}).
The exact status of these variables is still under debate. The intermediate position of 
RRd stars in the instability strip between RRc (pulsating in the first overtone mode) and RRab stars (pulsating in the fundamental mode) 
suggests that these stars are in the process of mode switching. This scenario seems however excluded by theoretical calculations 
(Cox, King \& Hodson \cite{ckh}) which yield a much too short duration for this transition state ($\lesssim10^{3-4}$ years) 
to account for the high fraction of RRd 
variables observed in some systems (e.g. IC4499, M68). It seems however that these stars are evolving rapidly (i.e. the changes 
are observable during a human life-time), and some period and amplitude changes have already been observed (Purdue et al. \cite{purdue}, 
Clement et al. \cite{chkr}, Papar\'o et al. \cite{pssas}, Benk\"o \& Jurcsik \cite{bj}). \\
\indent The first RRd star discovered was AQ Leo (Jerzykiewicz \& Wenzel \cite{jw}), a field RR Lyrae. After this discovery, searches were
focused on old stellar systems harboring RR Lyrae stars and, surprisingly, RRd variables were searched successfully in some systems 
but vainly in others. RRd stars have been found in globular clusters (M15: Nemec \cite{nemec2}, M3: Nemec \& Clement \cite{nc}, 
NGC2419 and NGC2466: Clement \& Nemec \cite{cns}, M68: Clement et al. \cite{cfs}, IC4499: Walker \& Nemec \cite{wn}), in Dsph 
galaxies (Draco: Nemec \cite{nemec1}, Sculptor: Ka\l u\.zny et al. \cite{kal}) and in the Galactic Halo (Clement et al. \cite{cks}, 
Garcia-Melendo \& Clement \cite{gc}, Clementini et al. \cite{ctf}). On the other hand, searches for RRd variables were unsuccessful 
in $\omega$Cen 
(Nemec et al. \cite{nnn}), M80, M9 and NGC2298 (Clement \& Walker \cite{cw}), Ursa Minor (Nemec et al. \cite{nwo}) and 20 other 
globular clusters (Clement \& Nemec \cite{cns}). The parameter(s) driving the occurrence of RRd pulsators in stellar systems is still not 
clear and more observations are needed before any firm conclusion can be drawn. What is clear, however, is that the two Oosterhoff 
groups are well separated in a Petersen diagram (a plot of the period ratio versus the fundamental period - Petersen \cite{peter}), 
the OoI RRd pulsators having lower periods and period ratios than their OoII counterparts. \\
\indent Recently, the collection of known RRd stars has been substantially increased with the discovery of 181 new RRd variables in 
the Large Magellanic Cloud (Alcock et al. \cite{macho97}, Alcock et al. \cite{macho}). These stars revealed a new picture because they 
were spread across the Petersen diagram, filling the gap between the two Oosterhoff groups. Pulsation models show that this distribution 
may be caused by a mass and/or metallicity spread within the population of RRd stars (Kov\'acs \cite{kovacs}). Spectroscopic measurements 
on a sample of these stars seems to confirm the metallicity spread (Clementini et al. \cite{cbdcg}, Bragaglia et al. \cite{bragaglia}). \\
\indent The search for multi-periodic RR Lyraes in the MACHO data set revealed other surprises. For instance, many RR Lyrae periodograms exhibited 
two closely spaced frequencies. This frequency pattern, first discovered by Olech et al. (\cite{olech1}, \cite{olech2}) in RR Lyrae stars, 
cannot be explained by the superposition of radial pulsations and is therefore believed to be related to non-radial modes. 
Although these kind of stars have been detected in only four different places to date (M55, M5, LMC and Galactic Bulge), they seem to be 
relatively common in their host system. Theoretical modelling of these stars have been proposed by Van Hoolst et al. (\cite{vh}) and 
by Dziembowski \& Cassisi (\cite{ds})\\
\indent We carried out a period analysis on a sample of $\sim$3\,700 RR Lyrae stars in Sgr and the Galactic Centre. 
The paper is structured as follows. Sect. 2 presents the data and their reduction. In Sect. 3, we describe the selection processes and 
present the sample of detected multi-periodic RR Lyraes. In Sect. 4, we present the period distributions of the RR Lyrae catalogs and 
compare them with those observed in other stellar systems. Sect. 5 is devoted to a presentation of the period-amplitude diagram of the RR Lyrae 
population in Sgr. In Sect. 6, we use photometric indicators to estimate the metallicity of the RR Lyrae population. Sect. 7 is devoted 
to a discussion about the spatial homogeneity of the RR Lyrae population and in Sect. 8 we estimate the RRab content of Sgr. 
Finally, we summarize our results and conclude in Sect. 9.
%
%
\section{Data}
\subsection{Observations and reduction}
\begin{figure}
 \resizebox{\hsize}{!}{\includegraphics{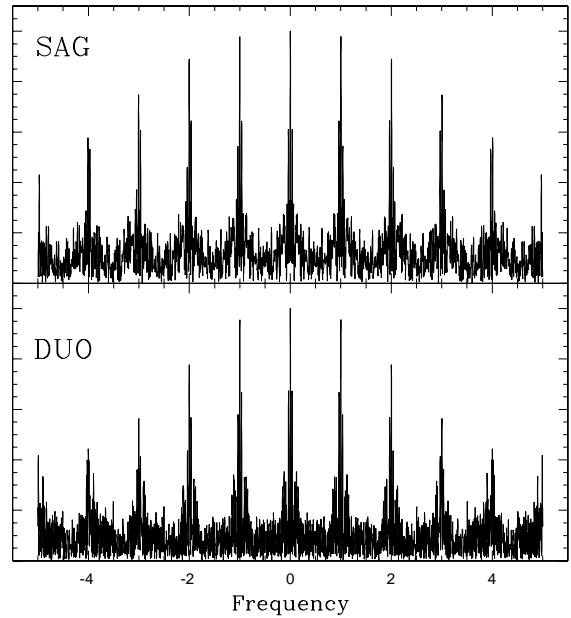}}
 \caption{Typical spectral windows for \textit{SAG} (Upper panel) and \textit{DUO} (lower panel).}
 \label{sw}
\end{figure}
\indent The data discussed in this paper are based on two sets of photographic plates taken with the 
ESO 1m Schmidt telescope at La Silla (Chile) between 1994 and 1996.
The first set of observations consists of a series of 82 B${_{\rm J}}$ photographic plates 
centred on (l,b)=(3.1$^{\circ}$,-7.1$^{\circ}$) and taken between June 11$^{\rm th}$ 1994 
and October 7$^{\rm th}$ 1994. The limiting magnitude of these plates reaches B$_{J}\sim$20.5. 
A first analysis of these data allowed the detection of microlensing 
events toward the Galactic bulge as part of the DUO project (Alard \& Guibert \cite{ag97}).
The second set of data results from a series of 68 Kodak Tech-Pan 4415 films combined with a BG12 filter 
centred on (l,b)=(6.6$^{\circ}$,-10.8$^{\circ}$). These observations spanned 83 days between 
May 17$^{\rm th}$ and August 9$^{\rm th}$ 1996. The exposure times resulted in a limiting magnitude of V$\sim$20. 
Together with the first set of plates, this series allowed us to reveal 
the shape of the Sagittarius dwarf galaxy near the Galactic Plane (Alard \cite{a96}; Cseresnjes, Alard \& Guibert \cite{cag}, hereafter 
Paper I). For convenience we will refer to the first field as the \textit{DUO} field whereas the second field 
will be called \textit{SAG} field. Typical spectral windows for \textit{SAG} and \textit{DUO} are shown in Fig.\ref{sw}.\\
\indent The plates were scanned at CAI/Paris Observatory with the high-speed 
microdensitometer MAMA\footnote{MAMA (http://dsmama.obspm.fr) is operated by INSU 
(Institut National des Sciences de l'Univers) and the Observatoire de Paris.}. The extraction of the sources
 were performed with the software \textit{Extractor} written by Alard (Alard \& Guibert \cite{ag97}). The final 
database contains the light curve for $\sim$20 10$^{6}$stars.
\subsection{Photometry}
\indent The B${_{\rm J}}$ band is well known to astronomers and it suffices here to recall the 
photometric relation linking this band to the standard Johnson system 
B${_{\rm J}}$=B-0.28(B-V) (Blair \& Gilmore \cite{bg}). On the other hand, 
the emulsion used in the second field may not be 
familiar to the reader so we shall say some words about it. 
The Tech-Pan 4415 emulsion is a fine grained, high resolution film 
with a sensitivity extending to 0.69$\mu$m. Together with a BG12 filter it results in a band peaking 
at $\sim$0.39$\mu$m with a width of $\sim$0.15$\mu$m at half maximum transmission (see Fig.1 in Paper I). 
We refer the interested reader to 
Phillipps \& Parker (\cite{pp}), Parker \& Malin (\cite{pm}) and the Kodak Web site\footnote 
{http://www.kodak.com/cluster/global/en/professional/ support/techPubs/p255/p255.shtml} for a more 
thorough description of the emulsion.
The band resulting from the combination 4415+BG12 has been calibrated with a sequence of 1638 stars 
located within our field and which were published 
by Sarajedini \& Layden (\cite{sl}). This calibration resulted in the relation B$_{4415}$=V+1.47(V-I). 
%
%
\section{Selection processes}
\subsection{Single-mode RR Lyraes}
The selection process for RR Lyrae stars is similar to the one used in Paper I, except that we adapted the 
parameters to allow detection of RRc stars. In short, we extracted from the data-base the variables with 
an amplitude $\gtrsim$0.2 mag. ($\chi^{2}/N_{DOF}>$5) and a minimum of 30 points in their light curve. 
These stars were then period searched between 0.2$^{\rm d}$ and 10.0$^{\rm d}$ using the multi-harmonic 
periodogram method of 
Schwarzenberg-Czerny (\cite{sc}) and we fitted a Fourier series with up to five harmonics to the folded 
light curve. The variables with a well-defined light curve ($\sqrt{\chi{2}/\chi^{2}_{fit}}>$1.5) were 
then plotted in the R$_{21}$/$\Phi_{21}$ plane where the RR Lyrae stars could easily be spotted. The final 
sample contains $\sim$3\,700 RR Lyrae variables almost evenly distributed between Sgr and the Galaxy.
\subsection{Multi-periodic RR Lyraes}
\subsubsection{First method}
\indent To search for multi-periodic RR Lyraes, we use the following procedure. For each lightcurve we search for the dominant period and 
fit a third order Fourier series to the folded lightcurve. The resulting Fourier series is then subtracted from the time series and the 
procedure is iterated. To characterize the strength of the peak value in the periodogram, we follow Alcock et al. (\cite{macho}) and calculate  
the statistics:
\begin{equation}
 S_{i}=\frac{A_{p}-\langle A(\nu)\rangle}{\sigma_{A(\nu)}}
\end{equation}
where $A_{p}$, $\langle A(\nu)\rangle$ and $\sigma_{A(\nu)}$ are the peak value, average value and standard deviation of the periodograms 
respectively. $S_{i}$ corresponds to the $i^{\rm th}$ cycle. The procedure is iterated as long as $S_{i}>8$. \\
\indent For all the lightcurves passing at least 2 iterations, we calculated the value $\chi_{ratio}$=$\chi^{2}_{1}$/$\chi^{2}_{2}$, 
where $\chi^{2}_{1}$ is the reduced $\chi^{2}$ of the fit of the Fourier series with the primary period and $\chi^{2}_{2}$ refers 
to the fit of the double Fourier series with the primary and secondary periods. All lightcurves for which Proba($\chi >\chi_{ratio})<50\%$ 
according to a Fisher-Snedecor distribution were selected for visual inspection. For all these candidates, we simultaneously checked 
 the lightcurves and the periodograms in order to select the multi-mode pulsating stars.
\subsubsection{Second method}
\begin{figure}
 \resizebox{\hsize}{!}{\includegraphics{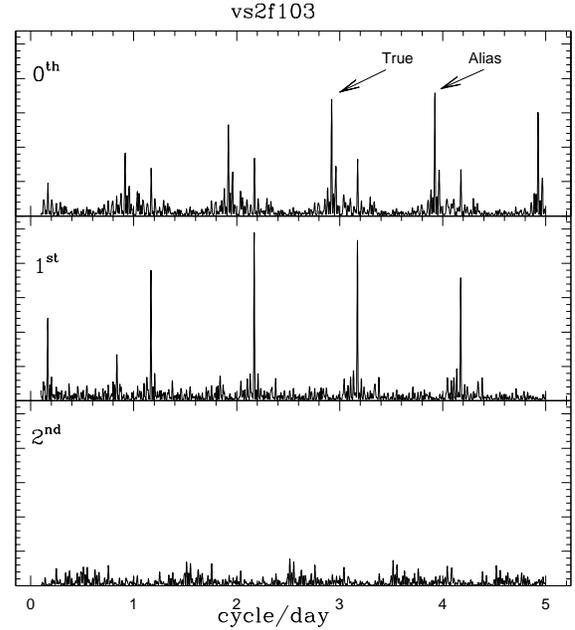}}
 \caption{Example of RRd star where the primary period is trapped in an alias period.}
 \label{alias}
\end{figure}
\indent It is very likely that we miss many multi-mode variables with the above-described procedure. This is due to three 
reasons: (1) the photometric accuracy is limited with our photographic material, 
(2) the superposition of two (or more) periods introduces noise in the periodograms, (3) the strong aliasing at $\pm n \, {\rm day}^{-1}$ 
(see Fig.\ref{sw}). There is not much that can be done about these problems, except for problem (3) in the search for RRd stars 
where we have more 
control because the period ratios are relatively well constrained. Fig.\ref{alias} shows a typical example of a RRd star missed by the 
preceding method. Although the maximum amplitude occurs at $\nu_{\rm max}=3.9232 {\rm d}^{-1}$, it is clear that the true frequency 
is rather $\nu_{0} =2.9232 {\rm d}^{-1}$, a value that yields a more ``classical'' period ratio $\nu_{0}/\nu_{1}=0.7425$.\\
\indent We thus re-processed the whole variable star data-base in order to search for double-mode RR Lyrae stars,  but this time  
we forced the period search between 0.33$^{\rm d}$ and 
0.44$^{\rm d}$. This interval encompasses the range of primary periods of all known RRds. A Fourier series 
with the periodicity P$_{1}$ was fitted to the light curve and the residual relative to this fit was 
period searched in the range 0.45$^{\rm d}<{\rm P}_{0}<$0.60$^{\rm d}$. The fit to the residuals was then 
subtracted from the initial light curve and the procedure was repeated. We then calculated the reduced 
$\chi^{2}$ about the resulting double Fourier series. \\
\indent In the following, we kept only those stars for which 0.74$\leqslant$P$_{1}$/P$_{0}$$\leqslant$0.75, 
corresponding to a range that is considered physically plausible by stellar pulsation models (Cox et al. \cite{ckh}, 
Kov\'acs et al. \cite{kbm}, Bono et al. \cite{bccm}).
To select the double mode pulsating RR Lyraes we define the variable  
$\chi_{ratio}$=$\chi^{2}_{1}$/$\chi^{2}_{2}$, where $\chi^{2}_{1}$ and $\chi^{2}_{2}$ are the reduced $\chi^{2}$ referring 
to the single-mode 
and double-mode fit respectively. All the variables for 
which the double-mode fit yields a better representation of the light curve (i.e. those with Proba($\chi >\chi_{ratio})<50\%$) 
where inspected 
visually. We simultaneously checked the single period and the double-mode fit in order not to spuriously select 
a variable with a true period outside the range searched as an RRd star. \\
\indent For completeness, we repeated the above-described 
procedure by inverting the order of the period searches (P$_{0}$ before P$_{1}$) in order to detect any eventual RRd with a higher 
amplitude in the fundamental mode relative to the first overtone mode. No additionnal RRd star was found this way, 
confirming that the first overtone mode is almost always the primary pulsation. This method allowed detection of 16 additional RRd stars.
\begin{table*}
 \caption{Double mode RR Lyrae in Sgr. Amplitudes are given in B$_{J}$ for \textit{DUO} and B$_{4415}$ for \textit{SAG}.}
 \begin{tabular*}{17.5cm}{l @{\extracolsep{\fill}} l @{ } l @{ } l @{ } l @{ } l @{ } l @{ } l @{ } l @{ } l @{ } l}
 \hline
 Ident. & $\alpha_{J2000}$ & $\delta_{J2000}$ & B$_{\rm J}$ & B$_{4415}$ & A$_{0}$ & A$_{1}$/A$_{0}$ & P$_{0}$ & P$_{1}$/P$_{0}$  & P$_{\rm F}$ & Method 1\\
 \hline \\
 vs3f773  & 18 31 53.88  & -28 49 07.1 & ...  & 19.1 & 0.16 & 2.33 & 0.60371 & 0.7447 & 0.00 & yes         \\
 vs10f77  & 18 53 24.68  & -29 43 46.2 & ...  & 19.1 & 0.17 & 3.23 & 0.46500 & 0.7432 & 0.07 & yes         \\
 vs4f75   & 18 53 39.03  & -29 19 05.9 & ...  & 19.0 & 0.29 & 1.83 & 0.55012 & 0.7466 & 0.06 & yes         \\
 vs2f99   & 18 52 26.75  & -28 24 57.1 & ...  & 19.1 & 0.41 & 1.34 & 0.47776 & 0.7435 & 0.03 & yes         \\
 vs2f103  & 18 52 36.04  & -29 15 37.4 & ...  & 19.3 & 0.46 & 1.15 & 0.46073 & 0.7425 & 0.06 & no (A1)     \\
 vs7f161  & 18 50 44.32  & -29 41 52.1 & ...  & 19.3 & 0.31 & 1.74 & 0.46502 & 0.7426 & 0.06 & no (A01)    \\
 vs1f242  & 18 48 31.65  & -29 00 15.1 & ...  & 19.2 & 0.39 & 1.49 & 0.52240 & 0.7460 & 0.01 & yes         \\
 vs1f243  & 18 48 06.14  & -29 12 44.6 & ...  & 19.1 & 0.35 & 1.43 & 0.45935 & 0.7423 & 0.08 & yes         \\
 vs17f249 & 18 48 23.98  & -30 25 14.0 & ...  & 19.1 & 0.41 & 1.29 & 0.47735 & 0.7440 & 0.02 & yes         \\
 vs9f385  & 18 43 29.01  & -29 43 01.8 & ...  & 18.9 & 0.20 & 2.55 & 0.47961 & 0.7442 & 0.09 & yes         \\
 vs2f443  & 18 41 59.37  & -30 06 23.6 & ...  & 19.1 & 0.33 & 1.54 & 0.46672 & 0.7432 & 0.02 & yes         \\
 vs1f524  & 18 39 49.07  & -29 29 30.2 & ...  & 19.3 & 0.27 & 1.78 & 0.54114 & 0.7461 & 0.05 & yes         \\
 vs14f598 & 18 36 34.95  & -27 36 40.6 & ...  & 19.7 & 0.35 & 1.49 & 0.54462 & 0.7456 & 0.07 & yes         \\
 vs3f659  & 18 34 53.71  & -28 32 48.5 & ...  & 19.7 & 0.29 & 1.97 & 0.47647 & 0.7430 & 0.05 & yes         \\
 vs8f105  & 18 52 44.12  & -29 39 24.1 & ...  & 19.1 & 0.16 & 3.19 & 0.47942 & 0.7439 & 0.20 & yes         \\
 vs2f266  & 18 47 44.99  & -28 18 48.1 & ...  & 19.6 & 0.21 & 2.38 & 0.46990 & 0.7436 & 0.11 & no (A0)     \\
 vs5f267  & 18 47 23.95  & -28 28 45.6 & ...  & 18.9 & 0.17 & 2.88 & 0.48187 & 0.7448 & 0.15 & yes         \\
 vs12f466 & 18 41 10.15  & -29 04 42.5 & ...  & 19.1 & 0.16 & 3.00 & 0.47537 & 0.7432 & 0.15 & no (A0)     \\
 vs13f158 & 18 50 36.45  & -29 06 58.0 & ...  & 19.0 & 0.10 & 4.47 & 0.47258 & 0.7438 & 0.26 & yes         \\
 vs3f170  & 18 49 42.00  & -25 56 57.1 & ...  & 20.2 & 0.25 & 2.70 & 0.48731 & 0.7455 & 0.29 & yes         \\
 vs5f408  & 18 43 08.95  & -28 40 02.6 & ...  & 19.1 & 0.13 & 4.12 & 0.47129 & 0.7432 & 0.28 & no (A0)     \\
 vs5f468  & 18 41 13.63  & -29 32 09.6 & ...  & 19.1 & 0.12 & 4.09 & 0.48087 & 0.7445 & 0.25 & no (A0)     \\
 vs0f684  & 18 34 44.91  & -27 51 38.1 & ...  & 19.0 & 0.23 & 1.58 & 0.46836 & 0.7435 & 0.23 & no (A01)    \\
 vs9f356  & 18 44 53.06  & -29 33 16.1 & ...  & 18.7 & 0.14 & 3.11 & 0.47159 & 0.7434 & 0.44 & no (A01)    \\
 vs0f525  & 18 39 53.39  & -29 34 51.2 & ...  & 19.2 & 0.07 & 7.04 & 0.45021 & 0.7417 & 0.43 & no (A0)     \\
 vd4f61   & 18 31 14.89  & -28 14 51.3 & 18.7 & ...  & 0.25 & 1.52 & 0.47203 & 0.7441 & 0.07 & yes         \\ 
 vd1f63   & 18 31 07.68  & -28 29 54.2 & 18.1 & ...  & 0.19 & 2.74 & 0.47793 & 0.7440 & 0.08 & yes         \\ 
 vd0f138  & 18 30 00.81  & -32 14 25.3 & 18.3 & ...  & 0.42 & 1.02 & 0.47218 & 0.7435 & 0.02 & yes         \\ 
 vd1f176  & 18 27 52.04  & -28 46 27.1 & 18.3 & ...  & 0.32 & 1.31 & 0.49083 & 0.7443 & 0.09 & yes         \\
 vd9f228  & 18 25 57.57  & -28 05 47.5 & 18.2 & ...  & 0.41 & 0.99 & 0.45725 & 0.7425 & 0.10 & no (A0)     \\
 vd2f263  & 18 25 22.74  & -29 24 41.7 & 18.3 & ...  & 0.42 & 1.34 & 0.47374 & 0.7434 & 0.02 & yes         \\
 vd16f471 & 18 18 33.38  & -31 37 54.3 & 18.2 & ...  & 0.21 & 2.77 & 0.47716 & 0.7435 & 0.07 & yes         \\
 vd4f101  & 18 30 05.41  & -30 24 10.0 & 18.0 & ...  & 0.13 & 4.16 & 0.47238 & 0.7432 & 0.16 & yes         \\
 vd9f185  & 18 27 48.51  & -30 32 11.1 & 18.0 & ...  & 0.20 & 2.18 & 0.54381 & 0.7459 & 0.12 & yes         \\
 vd14f208 & 18 26 29.60  & -29 37 52.0 & 18.2 & ...  & 0.43 & 1.46 & 0.46635 & 0.7433 & 0.14 & yes         \\
 vd8f687  & 18 11 29.88  & -30 00 23.6 & 18.1 & ...  & 0.11 & 2.93 & 0.55626 & 0.7464 & 0.29 & yes         \\
 vd1f141  & 18 48 41.74  & -27 24 38.6 & 17.6 & ...  & 0.30 & 1.91 & 0.47590 & 0.7443 & 0.03 & yes         \\
 vd5f152  & 18 28 43.59  & -29 31 55.5 & 17.9 & ...  & 0.37 & 1.68 & 0.46495 & 0.7432 & 0.05 & yes         \\
 vd11f353 & 18 22 35.34  & -30 31 52.3 & 17.9 & ...  & 0.25 & 1.47 & 0.47361 & 0.7441 & 0.13 & yes         \\
 vd21f689 & 18 11 22.76  & -30 33 12.7 & 17.6 & ...  & 0.09 & 4.38 & 0.51433 & 0.7453 & 0.25 & yes         \\
\hline
 \end{tabular*}
 \label{t1}
\end{table*}
\begin{table*}
 \caption{Same as Table \ref{t1} for Galactic RRd stars.}
 \begin{tabular*}{17.5cm}{l @{\extracolsep{\fill}} l @{ } l @{ } l @{ } l @{ } l @{ } l @{ } l @{ } l @{ } l @{ } l}
 \hline
 Ident. & $\alpha_{J2000}$ & $\delta_{J2000}$ & B$_{\rm J}$ & B$_{4415}$ & A$_{0}$ & A$_{1}$/A$_{0}$ & P$_{0}$ & P$_{1}$/P$_{0}$  & P$_{\rm F}$ & Method 1\\
 \hline \\
 vs7f693  & 18 33 56.78  & -29 39 11.2 & ...  & 17.0 & 0.26 & 1.77 & 0.45626 & 0.7437 & 0.04 & yes      \\
 vs1f148  & 18 51 06.00  & -27 15 02.9 & ...  & 17.1 & 0.20 & 2.44 & 0.54103 & 0.7462 & 0.24 & yes      \\
 vs13f417 & 18 42 37.13  & -30 23 37.3 & ...  & 17.9 & 0.29 & 1.87 & 0.48769 & 0.7447 & 0.26 & no (A01) \\
 vs14f440 & 18 41 49.51  & -29 31 39.7 & ...  & 16.0 & 0.11 & 5.15 & 0.48187 & 0.7447 & 0.32 & no (A0)  \\
 vs11f158 & 18 50 46.72  & -29 09 22.8 & ...  & 17.9 & 0.52 & 1.18 & 0.50367 & 0.7451 & 0.01 & yes      \\
 vs7f54   & 18 54 46.05  & -30 32 51.9 & ...  & 17.5 & 0.51 & 2.55 & 0.43574 & 0.7405 & 0.05 & yes      \\
 vd2f86   & 18 29 54.97  & -27 32 14.8 & 16.6 & ...  & 0.42 & 1.72 & 0.55026 & 0.7460 & 0.13 & yes      \\ 
 vd9f521  & 18 16 54.04  & -30 29 51.5 & 14.7 & ...  & 0.73 & 0.82 & 0.46157 & 0.7424 & 0.07 & no (A0)  \\
 vd21f652 & 18 12 22.92  & -28 49 34.4 & 16.8 & ...  & 0.40 & 1.36 & 0.47482 & 0.7434 & 0.06 & yes      \\
 vd15f747 & 18 09 53.11  & -30 50 56.3 & 16.6 & ...  & 0.42 & 1.49 & 0.47632 & 0.7444 & 0.02 & yes      \\
 vd20f737 & 18 09 58.97  & -28 52 25.5 & 16.3 & ...  & 0.18 & 2.05 & 0.57766 & 0.7456 & 0.30 & no (A0)  \\
 vd6f488  & 18 18 07.38  & -29 33 01.5 & 15.3 & ...  & 0.18 & 2.50 & 0.47560 & 0.7437 & 0.33 & no (N)   \\
 vd5f715  & 18 11 09.59  & -30 07 20.1 & 16.5 & ...  & 0.13 & 2.03 & 0.55489 & 0.7487 & 0.38 & no (N)   \\
\hline
 \end{tabular*}
 \label{t2}
\end{table*}
\begin{table*}
 \caption{Same as Table \ref{t1} for RR Lyrae stars in Sgr with two closely spaced frequencies. Subscripts $p$ and $s$ refer to the primary and secondary pulsations respectively.}
 \begin{tabular*}{17.5cm}{l @{\extracolsep{\fill}} l @{ } l @{ } l @{ } l @{ } l @{ } l @{ } l @{ } l @{ } l @{ } l}
 \hline
 Ident. & $\alpha_{J2000}$ & $\delta_{J2000}$ & B$_{\rm J}$ & B$_{4415}$ & A$_{p}$ & A$_{s}$/A$_{p}$ & P$_{p}$ & P$_{s}$/P$_{p}$  & P$_{\rm F}$ & type\\
 \hline \\
 vs2f274             & 18 47 37.05  & -29 50 50.7 & ...  & 19.0 & 0.28 & 0.71 & 0.29782 & 1.0283 & 0.01 & RRc   \\
 vs4f335             & 18 45 42.04  & -30 52 34.3 & ...  & 19.0 & 0.33 & 0.39 & 0.26455 & 0.9601 & 0.21 & RRc   \\
 vs4f391             & 18 44 09.83  & -30 48 31.5 & ...  & 19.3 & 0.26 & 0.75 & 0.29862 & 0.9710 & 0.05 & RRc   \\
 vd8f264             & 18 24 33.85  & -29 28 29.4 & 18.2 & ...  & 0.31 & 0.68 & 0.26556 & 0.9981 & 0.06 & RRc   \\
 vd8f661             & 18 12 39.07  & -30 28 13.9 & 18.9 & ...  & 0.36 & 0.47 & 0.26133 & 0.9770 & 0.17 & RRc   \\
\hline
 \end{tabular*}
 \label{t3}
\end{table*}
\begin{table*}
 \caption{Same as Table \ref{t3} for Galactic RR Lyraes.}
 \begin{tabular*}{17.5cm}{l @{\extracolsep{\fill}} l @{ } l @{ } l @{ } l @{ } l @{ } l @{ } l @{ } l @{ } l @{ } l}
 \hline 
 Ident. & $\alpha_{J2000}$ & $\delta_{J2000}$ & B$_{\rm J}$ & B$_{4415}$ & A$_{p}$ & A$_{s}$/A$_{p}$ & P$_{p}$ & P$_{s}$/P$_{p}$  & P$_{\rm F}$ & type\\
 \hline \\
 vs4f114             & 18 51 30.86  & -26 04 18.7 & ...  & 18.2 & 0.32 & 0.28 & 0.22685 & 1.0074 & 0.35 & RRc   \\
 vs3f146             & 18 50 44.22  & -26 49 39.2 & ...  & 18.1 & 0.47 & 0.45 & 0.26258 & 1.0151 & 0.00 & RRc   \\
 vd1f77              & 18 31 38.39  & -31 19 14.2 & 16.4 & ...  & 0.35 & 0.52 & 0.29817 & 0.9778 & 0.05 & RRc   \\
 vd25f234            & 18 25 31.48  & -29 07 17.1 & 13.9 & ...  & 0.33 & 0.67 & 0.26223 & 0.9824 & 0.18 & RRc   \\
 vd8f324             & 18 23 02.84  & -30 14 34.2 & 15.1 & ...  & 0.65 & 0.35 & 0.25743 & 1.0027 & 0.18 & RRc   \\
 vd25f670            & 18 12 00.53  & -32 13 51.9 & 15.2 & ...  & 0.44 & 0.59 & 0.28959 & 1.0193 & 0.12 & RRc   \\
 vd8f704             & 18 11 19.24  & -27 55 06.6 & 16.9 & ...  & 0.41 & 0.56 & 0.27036 & 0.9600 & 0.19 & RRc   \\
 vd9f754             & 18 09 44.70  & -32 08 32.0 & 16.8 & ...  & 0.28 & 0.82 & 0.28066 & 1.0263 & 0.19 & RRc   \\
\hline
 \end{tabular*}
 \label{t4}
\end{table*}
\begin{figure*}
 \resizebox{\hsize}{!}{\includegraphics{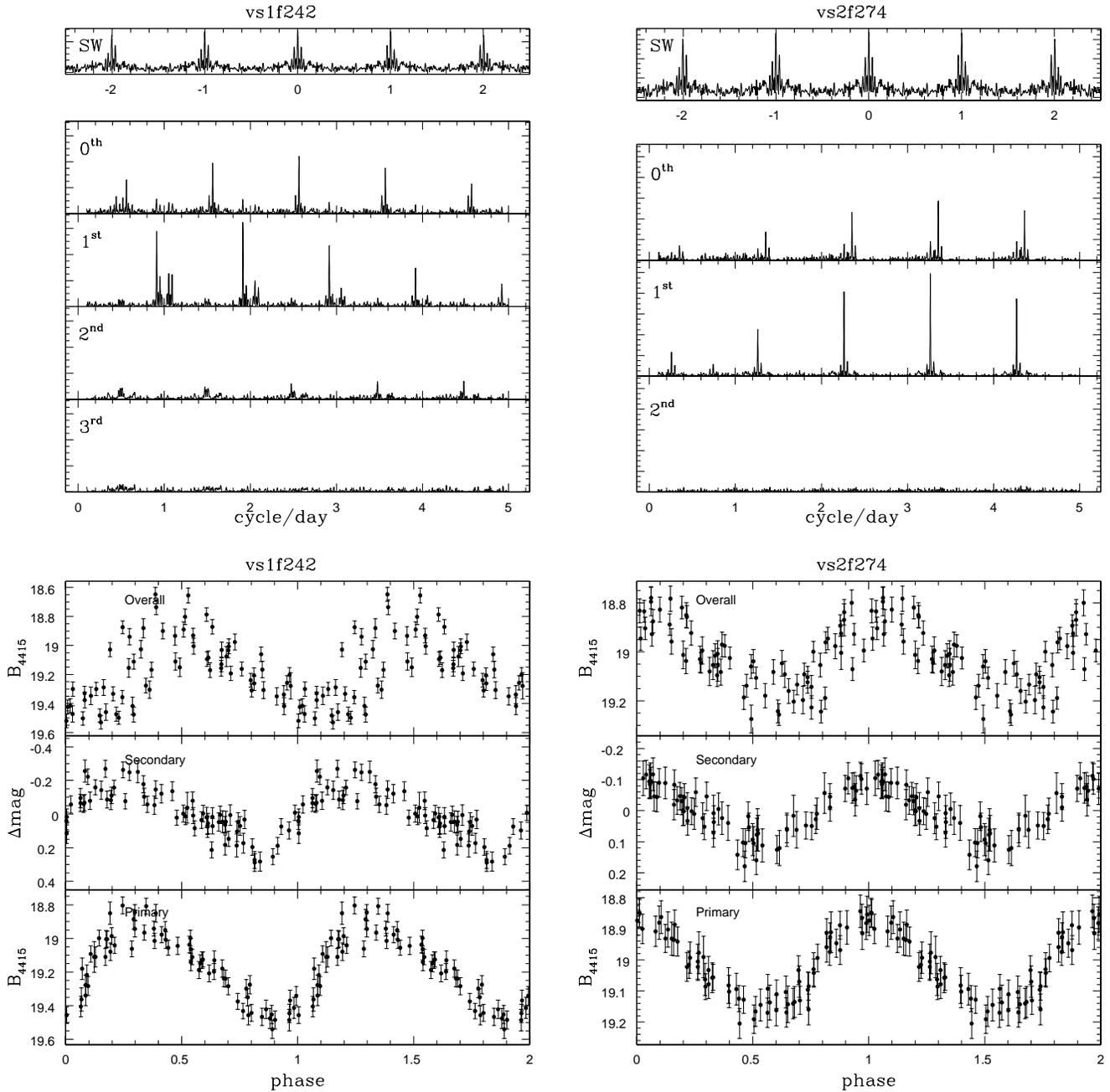}}
 \caption{{\bf Left panels:} a typical RRd star. The panels represent from top to bottom: spectral window, periodogram after successive subtraction of the main pulsation, and decomposition of the lightcurve. {\bf Right panels:} Same as before for a RR Lyrae star with two closely spaced frequencies.}
 \label{lc}
\end{figure*}
\subsubsection{Results}
\begin{figure*}
 \resizebox{\hsize}{!}{\includegraphics{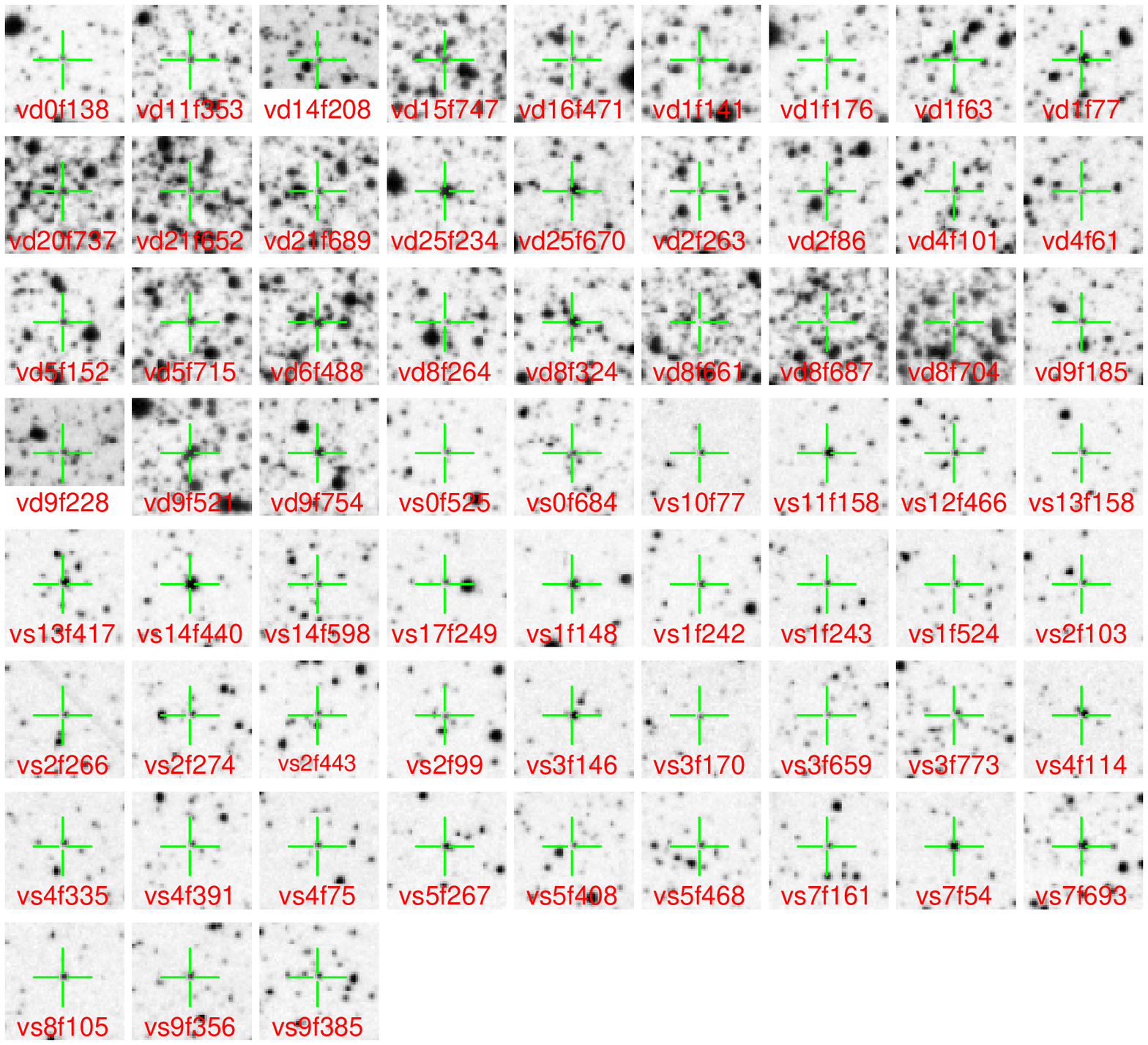}}
 \caption{Finding charts for multiperiodic RR Lyrae stars in Sgr and the Galactic Centre. Each finding chart is 36 \arcsec on a side. North is oriented up and East is left.}
 \label{find}
\end{figure*}
\indent The final sample contains two kinds of multiperiodic stars: RRd stars and RR Lyrae stars with two closely spaced frequencies. 
No attempt was made to 
detect RR Lyrae variables implying long time effects (e.g. Blazkho variables or period changes), because of our too short time span for 
these kinds of detections. Nor did we search for more than two periods because of the limited accuracy of our photometry.\\
\indent We found 53 RRd stars (40 in Sgr and 13 in the Galaxy), which are presented in 
Table \ref{t1} and \ref{t2}. The first 
column is the star identifier, columns 2 and 3 are the equatorial coordinates. The apparent magnitude of each star is shown in 
column 4 (\textit{DUO}) or 5 (\textit{SAG}). Columns 6 and 7 are the amplitude (in B$_{J}$ band for \textit{DUO} and 
B$_{4415}$ for \textit{SAG}) of the fundamental mode pulsation, and amplitude ratio respectively. 
Columns 8 and 9 are the periods and period ratios. Column 10 represents the probability that the 
$\chi_{ratio}$ occured by chance if it were following a Fisher-Snedecor distribution. Although this is not true because 
of the non-gaussianity of the errors, it is still indicative of the strength of the detection. In the last column, we indicate whether 
the RRd star has been detected through the first 
method or not, and in the latter case, why it was missed (A0: fundamental period aliased, A1: first overtone period aliased, N: 
periodogram too noisy - i.e. S$_{i}<8$). A typical RRd star is presented in Fig.\ref{lc} 
(left panels). \\
\indent In addition to the RRd stars, we found 13 RR Lyrae stars with two closely spaced frequencies (5 in Sgr and 8 in the Galaxy). 
These stars are presented in Table \ref{t3} (Sgr) and \ref{t4} (Galaxy). A typical detection is presented in Fig.\ref{lc} 
(right panels). Similar detections have been performed in M55 (Olech et al. \cite{olech1}), M5 (Olech et al. \cite{olech2}), 
the LMC (Alcock et al. \cite{macho}) and the Galactic Bulge (Moskalik \cite{moskalik}). \\
\indent Finding charts of all the detected multiperiodic RR Lyraes are presented in Fig.\ref{find}. North is up, East is left, and each 
box is 36\arcsec on a side. Light curves are available from the author upon request.
%
%
\section{Period distributions}
\subsection{Single-mode RR Lyrae stars}
\indent The mean periods of RR Lyrae stars in Sgr are 0.574$^{\rm d}\pm$0.002 and 0.322$^{\rm d}\pm$0.002 for RRab and RRc stars respectively. 
This can be compared to the mean periods found in OoI (0.55$^{\rm d}$ and 0.32$^{\rm d}$) and OoII (0.65$^{\rm d}$ and 
0.37$^{\rm d}$) groups. Sgr is thus to be classified in the OoI group. This is not common for Dsph galaxies, which are usually 
located in the intermediate region of the Oosterhoff dichotomy. 
\begin{table}
 \caption{Mean periods of RR Lyrae stars in dwarf galaxies of the local group. RR Lyrae data are taken from (Draco) Nemec (\cite{nemec1}), (Ursa Minor) Nemec et al. (\cite{nwo}), (Carina) Saha et al. (\cite{sms}), (Leo II) Siegel \& Majewski (\cite{sm}), (Sculptor) Ka\l u\.zny et al. (\cite{kal}), (Sextans) Mateo et al. (\cite{mfk}), (SMC) Smith et al. (\cite{ssbg}) and Graham (\cite{graham}), Caretta et al. (\cite{ccfft}), Silbermann \& Smith (\cite{ss}).}
 \begin{tabular}{l @{\extracolsep{0.05cm}}  c @{   }  c @{   }  c @{   }  c @{   }  c @{   }  c @{   }  c}
  \hline
  System            & [Fe/H] & $\langle P_{ab}\rangle$ & $\sigma_{ab}$ & $\langle P_{c}\rangle$ & $\sigma_{c}$ & $r$ & $p_{s}$ \\
  \hline
  M3                & -1.7               & 0.558 & 0.008 & 0.344 & 0.019 & 0.87 & -24.0   \\
  M15               & -2.2               & 0.641 & 0.013 & 0.359 & 0.008 & 0.15 & -1068.5 \\
  Draco             & -2.0               & 0.614 & 0.004 & 0.351 & 0.012 & 0.67 & -48.9   \\
  Ursa Minor        & -2.2               & 0.638 & 0.009 & 0.375 & 0.011 & 0.16 & -63.0   \\
  Carina            & -2.0               & 0.620 & 0.006 & 0.366 & 0.015 & 0.51 & -60.8   \\
  Leo II            & -1.9               & 0.619 & 0.006 & 0.363 & 0.008 & 0.62 & -42.9   \\
  Sculptor          & -1.8               & 0.587 & 0.007 & 0.337 & 0.005 & 0.67 & -46.7   \\
  Sextans           & -1.7               & 0.606 & 0.010 & 0.355 & 0.024 & 0.78 & -44.6   \\
  SMC (RRLyr)       & -1.7               & 0.588 & 0.006 & 0.380 & 0.008 & 0.81 & -34.5   \\
  LMC (RRLyr)       & -1.6               & 0.582 & 0.001 & 0.325 & 0.001 & 0.96 & -19.0   \\
  Sgr               & -1.6               & 0.574 & 0.002 & 0.322 & 0.002 & 1.00 & ...     \\
  Galactic Centre   & ...                & 0.548 & 0.002 & 0.305 & 0.002 & 0.86 & -28.3   \\
  \hline
 \end{tabular}
 \label{rr_period}
\end{table}
Table \ref{rr_period} presents the mean periods of RR Lyrae stars in an OoI type globular cluster (M3), an OoII type cluster (M15), 
in the Galactic Centre and in 
all the Milky Way satellites with a well-studied RR Lyrae population. The LMC RR Lyrae stars have been extracted from photographic plate 
time series of the first EROS season whereas sources for all the other systems are listed in the caption of Table \ref{rr_period}.\\
\indent  The RR Lyrae stars in Sgr present the shortest average periods among all the 
dwarf galaxies, but slightly longer than the OoI globular cluster M3 and the Galactic Centre. 
Indeed, these values are rather close to those found in the Large Magellanic Cloud ($\langle P_{ab}\rangle$=0.582$^{\rm d}\pm$0.001 
and $\langle P_{c}\rangle$=0.325$^{\rm d}\pm$0.001). 
\begin{figure}
 \resizebox{\hsize}{!}{\includegraphics{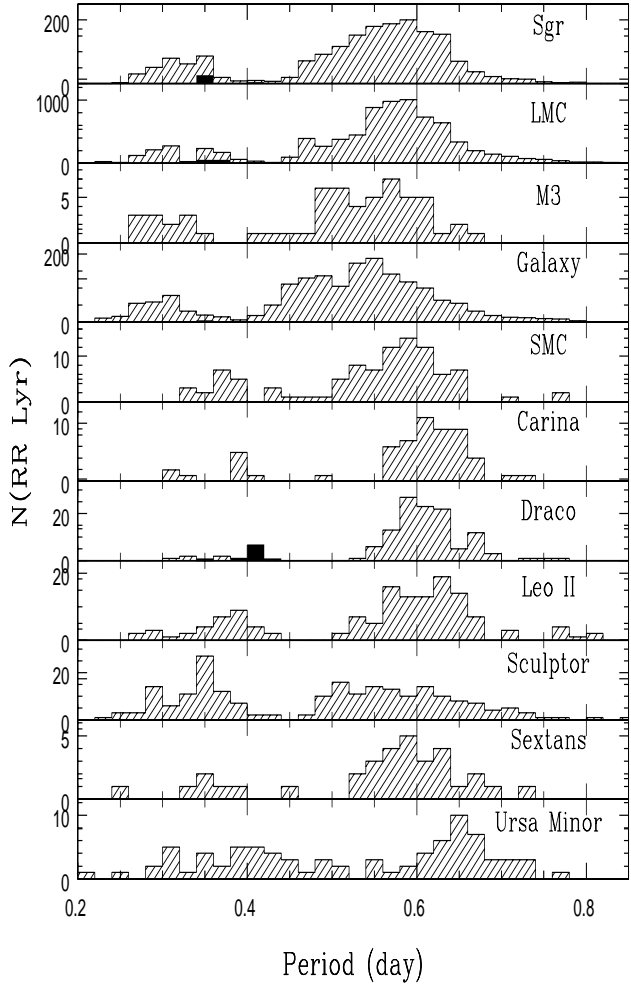}}
 \caption{Period distribution of the RR Lyraes detected in various systems. The dark regions indicate the location of the RRd variables 
(1st overtone period).}
 \label{period}
\end{figure}
\begin{figure}
 \resizebox{\hsize}{!}{\includegraphics{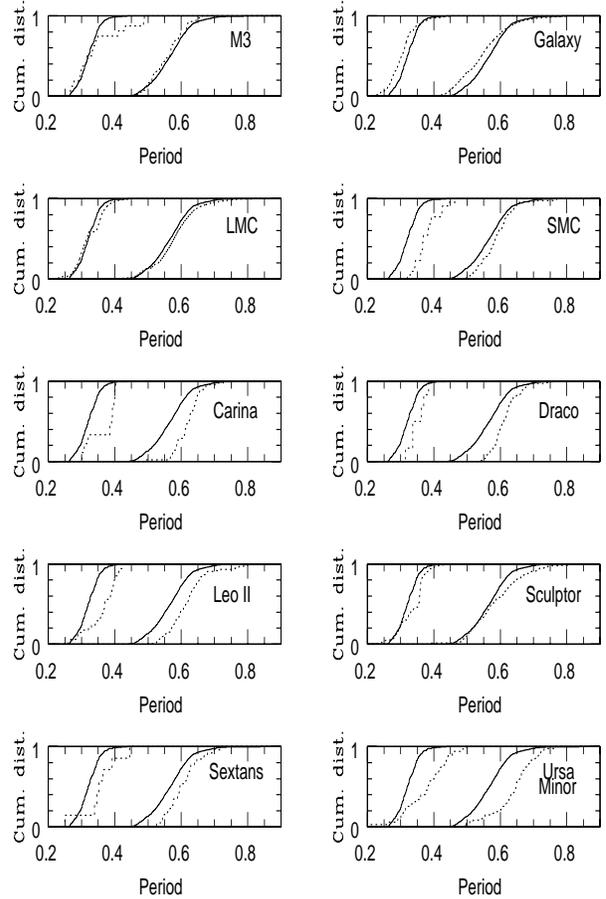}}
 \caption{Cumulative distributions of RR Lyrae periods in the OoI globular cluster M3 and all satellites of the Milky Way with a well studied RR Lyrae population. The solid line refers to Sgr whereas the dotted line corresponds to the system indicated in each panel. Left (resp. right) curves correspond to RRc stars (resp. RRab stars).}
 \label{cum_period}
\end{figure}
The similarity between Sgr and the LMC is even more striking when one considers the period distributions of the RR Lyraes.
The period distributions of RR Lyrae stars in the Galactic Centre and in different dwarf galaxies are shown in Fig.\ref{period} 
and are compared to that of Sgr in Fig.\ref{cum_period}. 
The coefficients of correlation between all these histograms and the period distribution in Sgr 
are shown in column 7 of Table \ref{rr_period}. As expected, the highest correlation is reached 
for the LMC\footnote{For the correlation between Sgr and the LMC we excluded the bins P $\in [0.32^{\rm d},0.34^{\rm d}]$ and 
P $\in [0.48^{\rm d},0.50^{\rm d}]$ because of poor phase coverage for the EROS stars in these period ranges caused by the nearly daily 
sampling of the light curves} where $r=0.96$. \\
\indent Since the correlation coefficient $r$ measures the linearity between two distributions, it is sensitive to the shape of 
the distributions, but not to \emph{shifts} (i.e. two identical distributions shifted one with respect to the other would have $r=1.00$). 
We thus performed another test, adapted from maximum likelihood statistics, and which is more sensitive to shifts. 
First, each period histogram is normalized to that of Sgr. We then calculate 
the values $p_{s}=\Sigma_{i}{\rm log}\,p_{i}$ where $p_{i}$ corresponds to the probability of finding the number of stars observed in the 
$i^{\rm th}$ bin if the parent distribution was identical to the period histogram of Sgr. The results are shown in column 8 of 
Table \ref{rr_period}. The highest $p_{s}$ value is reached by the LMC, confirming that the period distributions in Sgr and the LMC are 
the most similar. \\
\indent The RR Lyrae samples in Sgr and the LMC are large ($\sim$1\,700 in Sgr and $\sim$6\,500 in the LMC), 
making the resemblance between these two systems significant. This resemblance is even more striking when one considers that no two other 
dwarf galaxies have similar period distributions. Note however that the two distributions are not identical, this 
being excluded at the 99$\%$ level by a KS test. 
\subsection{Double-mode RR Lyrae stars}
\begin{figure}
 \resizebox{\hsize}{!}{\includegraphics{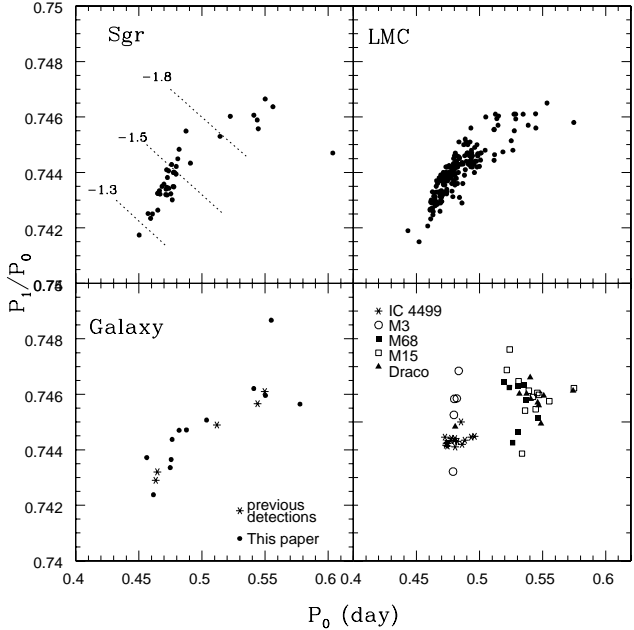}}
 \caption{Petersen diagram for all the published RRd. In the upper left panel, 
the dotted lines correspond to the metallicity for a specific model of RRd star (see text). From top to bottom [Fe/H]=-1.8, -1.5 and -1.3 dex. In the lower left panel, asterisks are RRd stars presented in the literature whereas heavy dots correspond to RRd stars presented in this paper.}
 \label{petersen}
\end{figure}
\begin{table}
 \caption{Summary of RRd stars detected to date. Metallicities are taken from Mateo (\cite{mateo}) for dwarf galaxies and from Harris (\cite{harris}) for globular clusters. Informations on RRd stars are taken from IC4499: Walker \& Nemec (\cite{wn}), M3: Nemec \& Clement (\cite{nc}), Corwin et al. (\cite{cca}), NGC2419 and NGC6426: Clement \& Nemec (\cite{cns}), M68: Clement et al. (\cite{cfs}), M15: Nemec (\cite{nemec2}), Draco Dsph: Nemec (\cite{nemec1}), Sculptor: Ka\l u\.zny et al. (\cite{kal}), LMC: Alcock et al. (\cite{macho}), MW: Clement et al. (\cite{cks}), 
Garcia-Melendo \& Clement (\cite{gc}), Clementini et al. (\cite{ctf})}
 \begin{tabular}{l @{\extracolsep{0.10cm}}  c @{   }  c @{   }  c @{   }  c @{   }  c}
  \hline
  System            &  [Fe/H]               & N(RRd)  & $\langle P^{RR_{d}}_{0}\rangle$ & $\langle P^{RR_{d}}_{1}\rangle$ & r \\
  \hline
  IC4499            &  -1.5                & 17      & 0.481 & 0.358 & 0.44 \\
  M3                &  -1.7                & 5       & 0.481 & 0.359 & 0.22 \\
  NGC 2419          &  -2.1                & 1       & 0.546 & 0.407 & ...  \\
  M68               &  -2.1                & 8       & 0.531 & 0.396 & -0.02\\
  M15               &  -2.2                & 12      & 0.541 & 0.404 & 0.13 \\
  NGC 6426          &  -2.2                & 1       & 0.542 & 0.404 & ...  \\
  Draco Dsph        &  -2.0                & 10      & 0.540 & 0.403 & 0.15 \\
  Sculptor          &  -1.8                & 1       &  ?    & 0.404 & ...  \\
  LMC               &  -1.7                & 181     & 0.485 & 0.361 & 0.77 \\
  Sgr               &  -1.1                & 40      & 0.487 & 0.362 & 1.00 \\
  MW                &  ...                 & 18      & 0.504 & 0.376 & 0.54 \\
  \hline
 \end{tabular}
 \label{system_rrd}
\end{table}
\indent Table \ref{system_rrd} summarizes the average periods of RRd stars in all the systems known to harbor this kind of variable. 
Again, the LMC and Sgr show a remarkable similarity in the mean periods of their RRd star population. In Fig.\ref{petersen} we present 
the distributions of all the published RRd stars in the Petersen diagram. The cluster RRd variables 
and those of the Draco Dsph galaxy occupy two distinct regions, reproducing the Oosterhoff dichotomy. This is no more the case for RRd stars 
in Sgr, in the LMC and in the Galactic Centre. These latter stars are distributed on a strip across the diagram. These distributions can be 
explained either by a metallicity spread, a mass spread or a combination of both (Kov\'acs \cite{kovacs}). 
A metallicity spread in the RRd population of the LMC has already been confirmed by spectroscopic measurements 
(Clementini et al. \cite{cbdcg}, Bragaglia et al. \cite{bragaglia}) and it is likely that the same applies to Sgr.  \\
\indent  To compare further, we 
transformed the Petersen diagrams into density maps and calculated the correlation between all these distributions. The results 
are shown in column 6 of Table \ref{system_rrd}. The correlation between the LMC and Sgr is $r$=0.77, which is not as good as 
the correlation between the 
single-mode RR Lyrae period distributions, but differences in the completeness between samples may affect the comparison.  
\section{Period-Amplitude diagram}
\begin{figure}
 \resizebox{\hsize}{!}{\includegraphics{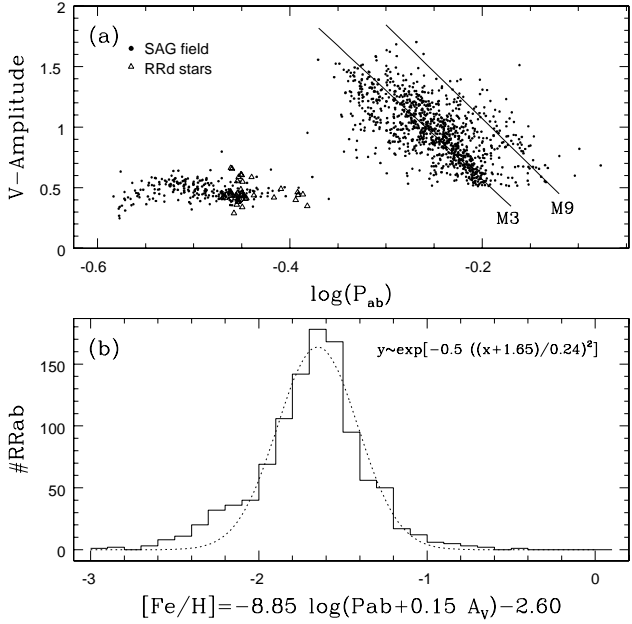}}
 \caption{{\bf Panel a:} period/amplitude diagram of the RR Lyrae stars in Sgr. Dots represent stars detected in \textit{SAG} and triangle are double-mode RR Lyraes. Also shown are the fiducial lines of M3 (OoI group) and M9 (OoII group). {\bf Panel b:} Metallicity estimate from the period-amplitude-metallicity relation. The dotted line represents the best Gaussian fit, whose relation is given in the upper right corner of the panel.}
 \label{per_amp}
\end{figure}
\indent The period-amplitude diagram is presented in Fig.\ref{per_amp}a. To construct this figure, we considered only the \textit{SAG} 
field, which has a more precise photometry than \textit{DUO}, resulting in a higher accuracy of the amplitudes. As will be shown in 
Sect. 7, the RR Lyrae population is homogeneous over the field and there is therefore no loss of generality by restricting 
ourself to \textit{SAG}. We rescaled the amplitudes in \textit{SAG} to the V amplitude 
from the relation A$_{\rm V}=0.93(\pm 0.17){\rm A}_{4415}-0.05(\pm 0.16)$, inferred from 15 RR Lyrae stars in common with 
those detected by Layden \& Sarajedini (\cite{ls}) near M54. \\
\indent The distribution of RRab stars in Fig.\ref{per_amp}a presents a high dispersion but is preferentially clumped 
around the ridge line of the globular cluster M3, confirming that this population is related to the OoI group. However, the 
distribution is skewed toward the ridge line of the OoII cluster member M9, suggesting the presence of a metal-poor sub-population 
within the RR Lyrae population. The RRab population is dominated by low-amplitude stars with a ratio 
$N(A_{V}>1.0{\rm mag})/N(A_{V}<{\rm 1.0mag})$=0.55. 
The RR Lyrae population is thus dominated by RRb stars, indicative of a red horizontal branch morphology. \\
\indent Remarkably, the same features are apparent in the period-amplitude diagram of the 
LMC (Alcock et al. \cite{alc96}, Alcock et al. \cite{alc00}), providing new evidence of the similarity of the RR Lyrae populations 
between these two galaxies.
\section{Metallicity estimate of the RR Lyrae population}
\subsection{Period-Metallicity relations}
\indent Some indication on the metallicity of the RR Lyrae population in Sgr can be inferred from its period distribution. 
 In a study of cluster and field RR Lyrae stars spanning a wide range of metallicities, Sandage (\cite{s93}) 
related the average periods of RR Lyrae stars to their metallicities:
\begin{equation}
 {\rm [Fe/H]_{ZW}}=(-{\rm log}\langle P_{ab}\rangle-0.389)/0.092
 \label{e1}
\end{equation}
\begin{equation}
 {\rm [Fe/H]_{ZW}}=(-{\rm log}\langle P_{c}\rangle-0.670)/0.119
 \label{e2}
\end{equation}
where $\langle P_{ab}\rangle$ and $\langle P_{c}\rangle$ are the average periods of RRab and RRc stars respectively. 
Although these relations were derived from cluster and field RR Lyraes, Siegel \& Majewski (\cite{sm}, their Fig.6) showed that 
RR Lyrae populations in Dsph followed the same relation. Applying Eqs.\ref{e1} and \ref{e2} to Sgr yields an average metallicity of 
[Fe/H]=$-$1.61 and $-$1.49dex respectively.\\
\indent The location of the blue and red fundamental edge of the instability strip are functions of metallicity. The shortest and longest 
period RR Lyraes are thus indicative of the metallicity boundaries of the RR Lyrae population which are given as (Sandage \cite{s93}):
\begin{equation}
 {\rm [Fe/H]_{ZW}}^{\rm (max)}=[-{\rm log}(P^{min}_{ab})-0.500]/0.122-0.2
 \label{e3}
\end{equation}
\begin{equation}
 {\rm [Fe/H]_{ZW}}^{\rm (min)}=[-{\rm log}(P^{max}_{ab})-0.280]/0.090-0.2
 \label{e4}
\end{equation}
where $P^{min}_{ab}$ 
and $P^{max}_{ab}$ are the minimum and maximum periods of RRab stars respectively. We added the constant terms in Eqs.\ref{e3} and \ref{e4} 
in order to rescale the metallicity to the Zinn \& West (\cite{zw}) scale, whose zero point is $\sim$0.2dex more metal-poor than the 
Butler \& Blanco metallicity scale (Blanco \cite{blanco}) on which these equations are based. 
 The shortest RRab period is 0.41531$^{\rm d}$,  
yielding an upper limit of [Fe/H]=$-$1.17dex,whereas the longest period of 0.84400$^{\rm d}$ implies a lower limit of [Fe/H]=$-$2.49dex. These 
values suggest a considerable spread in the metallicity of the RR Lyrae population. 
\subsection{Period-Amplitude-Metallicity relation}
\indent  Alcock et al. (\cite{alc00}) provide a period-amplitude-metallicity relation calibrated with high-quality V-band light curves 
of RR Lyrae stars in several globular clusters:
\begin{equation}
 {\rm [Fe/H]}=-8.85\,{\rm log}(P_{ab}+0.15\,A_{V})-2.60
 \label{e5}
\end{equation}
where $A_{V}$ is the amplitude in the V-band. The resulting metallicity distribution of Sgr RRab member stars is presented in 
Fig.\ref{per_amp}b. The peak-value is at 
[Fe/H]$\simeq -$1.65dex, close to the estimate inferred from the Sandage relations. This is also similar to the value found in the LMC 
with the same relation ([Fe/H]$\simeq -$1.6; Alcock et al. \cite{alc00}). \\ 
\indent The best Gaussian fit to the metallicity distribution is given by $y\propto e^{-\frac{1}{2}\big(\frac{x+1.65}{0.24}\big)^{2}}$. As 
can be seen in Fig.\ref{per_amp}b, this function fits the distribution relatively well, except for ${\rm [Fe/H]} \lesssim -$2.0dex where the presence 
of a significant metal-poor subpopulation is apparent.
\subsection{Petersen diagram}
\indent The strong metallicity spread in the RR Lyrae population seems confirmed by the distribution of 
RRd stars in the Petersen diagram, although a dispersion in the star masses could also be responsible for the spread. 
The dotted lines in Fig.\ref{petersen} represent metallicities for a specific model of RRd with a mass of 
0.75 M$_{\odot}$, a luminosity log($L/L_{\odot}$)=1.72, a hydrogen abundance X=0.76 and an effective temperature of 6900, 6800 and 
6700K for [Fe/H]=1.3, 
-1.5 
and -1.8 dex respectively. This model has been taken from Kov\'acs (\cite{kovacs}). The positions of the RRd stars are in good agreement 
with the above 
estimate of mean metallicity and metallicity spread of the RR Lyrae population. \\
\indent A clump of RRd stars is apparent in Fig.\ref{petersen} at ${\rm [Fe/H]} < -$1.8dex, suggesting the presence of a minor but significant 
population of low metallicity and/or high mass RR Lyraes. The loci of the long period RRd stars in the 
Petersen diagram is similar to those found in OoII systems. Table \ref{system_rrd} summarizes all the systems with known RRd pulsators. 
One sees that 
all these OoII systems have a metallicity within $-$2.0 and $-$2.2dex, supporting the assumption that a fraction of RR Lyrae stars in Sgr has 
this abundance.\\
\indent The existence of a very low metallicity population in Sgr has been 
suggested by Bellazzini et al. 
(\cite{bel1}, \cite{bel2}) who detected a star count excess in a region of the CMD that could represent a very blue horizontal branch. 
Furthermore, in a period-amplitude diagram of RR Lyrae stars towards the globular cluster M54, Layden \& Sarajedini (\cite{ls}) noted 
that a fraction of RR Lyrae stars could be consistent with 
a contamination by Sgr field stars of metallicity ${\rm [Fe/H]}=-2.1$dex.
\section{Homogeneity of the RR Lyrae population over the field}
\indent Several authors claim having detected a metallicity gradient in the stellar population of Sgr, the centre being more 
metal-rich than the outer regions. This finding is 
based on the redder morphology of the Red Giant Branch and the Horizontal Branch at the centre of Sgr relative to the outer regions 
(Marconi et al. \cite{marconi}, Bellazzini et al. \cite{bel2}, Alard \cite{a01}). If confirmed, this feature would 
be in concordance with observations of other dwarf galaxies (e.g. Saviane et al. \cite{shmr} and references therein).
\begin{figure}
 \resizebox{\hsize}{!}{\includegraphics{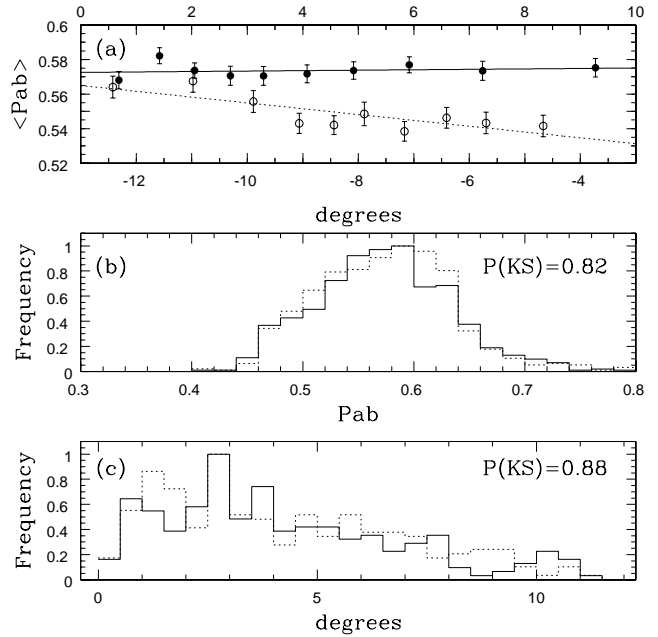}}
 \caption{{\bf Panel a:} average period of RRab stars as a function of position. Filled circles correspond to RRab variables in Sgr and a distance (X) from M54 projected onto the main axis (upper labels). Open circles are average periods of Galactic RRab and are plotted as a function of Galactic latitude (lower labels). {\bf Panel b:} Normalized period histograms of RRab stars. The solid line correspond to RRab stars with X$<$5$^{\circ}$ whereas the dotted line correspond to RRabs with X$>$5$^{\circ}$. {\bf Panel c:} Normalized histograms of X for the 250 shortest period (solid line) and longest period (dotted line) RRab variables in Sgr.}
 \label{homogeneity_rrab}
\end{figure}
We now question whether such a gradient is apparent in the RR Lyrae population. \\
\indent Any metallicity gradient would translate into a gradient in the mean period of RRab stars.
Fig.\ref{homogeneity_rrab}a presents the mean period of RRab stars in Sgr (solid line) as a function of distance from M54 projected onto the 
main axis\footnote{We assume a position angle of 108.6$^{\circ}$ for Sgr.}. Each bin size has been adapted in order to contain the 
same number of RRab stars. A linear least square fit through 
these points yields a slope of $\partial (\langle P_{ab}\rangle)/\partial X=0.0002(\pm0.0004)$. From Eq.\ref{e1}, 
it results in a metallicity difference $<$0.1dex between 
the two extremities of the field. Clearly, there is no significant metallicity gradient in the RR Lyrae 
population of Sgr. For comparison, we present in Fig.\ref{homogeneity_rrab}a the average period of Galactic RRab as a function of 
latitude (dotted line), where the period dependence as a function of position is evident. \\
\indent To test the homogeneity further, we divided the RRab catalogue into two subsamples around the median distance from M54 
and compared the period distributions (Fig.\ref{homogeneity_rrab}b). A KS test shows that the two histograms are drawn from the same 
parent distribution with a probability of 82$\%$, suggesting a similar horizontal branch morphology in the two subsamples. \\
\indent Finally, in Fig.\ref{homogeneity_rrab}c we compare the spatial distributions of the 250 
longest and shortest period RRab variables. The KS test yields a probability of 88$\%$ for the two histograms to be issued from the 
same parent distribution. This definitely excludes any modification of the period distribution with position. Thus, unless a metallicity gradient 
and the second parameter effect conspire to keep the 
horizontal branch morphology constant over the field, the RR Lyrae population should be homogeneous in the main body of Sgr. Furthermore, if 
the metallicity gradient was confirmed, this 
may indicate that the RR Lyrae population is not associated to the prominent red horizontal branch which is apparent in the CMDs of Sgr.\\
\section{Estimation of the number of RR Lyrae stars in Sgr}
\begin{figure}
 \resizebox{\hsize}{!}{\includegraphics{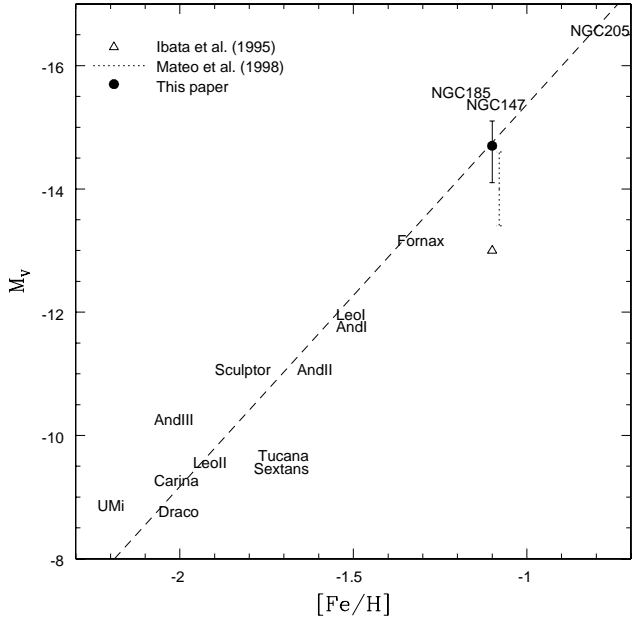}}
 \caption{Luminosity-metallicity relation for Dsph of the Local group. Data have been taken from Mateo (\cite{mateo}). The dashed line corresponds to the best fit to the data points. The filled circle corresponds to the estimated luminosity of Sgr if this galaxy had lost 50$\%$ of its mass since its formation. The open triangle represents the first luminosity estimate of this galaxy (Ibata et al \cite{igi}). The vertical dotted line (shifted by 0.02dex for clarity) encompasses the different estimates of Mateo et al. (\cite{mom}).}
 \label{mvdsph}
\end{figure}
\indent Using the results from paper I, it is possible to estimate the total number of RRab stars in Sgr. The best fit to the core of Sgr 
corresponds to an exponential with a scale-length of 4.1$^{\circ}$ on the main axis and a central density of 139 RRab per square degree. 
Assuming an axis ratio of 3:1 (Ibata et al. \cite{igi}), we get $\sim$4\,200 RRab in the main body of Sgr. This estimate is only a lower 
limit for Sgr since we did not consider the nearly flat profile in the outer region of the main body (Paper I). Irwin (\cite{irwin}) estimated 
that Sgr lost 50$\%$ of its mass through tidal stripping. This implies that the number of RRab stars associated with Sgr could be 
as high as $\sim$8\,400. \\
\indent Provided RRab stars are good 
tracers of light in Sgr, we can estimate a lower limit to the total luminosity of this galaxy using the luminosity 
function (LF) published by Mateo et al. (\cite{muskkk}). Their LF is complete down to M$_{\rm V}$$\sim$4.5 mag. For fainter 
magnitudes we extend the LF with the Bulge LF (Holtzman et al. \cite{holtz}). This should not significantly alter the result since most 
of the light is emitted by brighter stars. The field of Mateo et al. contains five RRab stars. Rescaling the RRab number density to the 
integrated luminosity and assuming a distance modulus of (m-M)$_{0}$=17.0 to Sgr yields $-$4.9 mag per RRab. It results that 
M$_{\rm V}{\rm (Sgr)}< -$13.9$^{+0.4}_{-0.6}$ mag, where the uncertainties represent the poissonian uncertainty in the RRab counts. Again, 
if we assume 50$\%$ mass loss, the total magnitude of Sgr before stripping would be $\sim -$14.7mag. \\
\indent Previous estimates have continuously increased the luminosity of Sgr as new extensions of this galaxy were discovered, with estimations 
ranging from M$_{\rm V}\sim -$13.0 (Ibata et al. \cite{igi}) to M$_{\rm V}\sim -$14.6mag (Mateo et al. \cite{mom}). Our estimation seems 
to favor the higher luminosity of Sgr. The metallicity of the dominant population in Sgr is estimated to be [Fe/H]$\sim -$1.1dex (Mateo et 
al. \cite{muskkk}). An absolute luminosity of $\sim -$14.7 mag for Sgr would be consistent with the empirical magnitude-metallicity 
relationship for Dsph galaxies, as shown in Fig.\ref{mvdsph}.\\
\section{Summary and conclusion}
\indent We summarize our results below:
\begin{enumerate}
 \item The average period of RR Lyrae stars in Sgr places this galaxy in the long period tail of the Oosterhoff I group.\\
 \item We found 53 double-mode RR Lyraes (40 in Sgr and 13 in the Galaxy), and 13 RR Lyrae stars with two closely spaced 
pulsation frequencies (5 in Sgr and 8 in the Galaxy). The multi-periodic RR Lyraes in Sgr are the first such stars ever discovered in this 
galaxy whereas the 13 foreground RRds increase the number of known Galactic RRd stars to 18.\\
 \item The period and amplitude distributions of the RR Lyrae population in Sgr suggest an average metallicity of $\sim$$-$1.6dex, with 
a contribution of a minor but perceptible population at ${\rm [Fe/H]}\lesssim$$-$2.0dex. Furthermore, together with the period distribution 
of RRd stars, there seems to be a large metallicity spread within the RR Lyrae population.\\
 \item We find no significant variation of the period distribution over our field, suggesting a homogeneous population. There is therefore 
no evidence of a metallicity gradient in the RR Lyrae population. This result is in 
contradiction with claims of a metallicity gradient within the main body of Sgr, unless the RR Lyrae population is not directly related 
to the other populations.\\
 \item We find a striking similarity between the RR Lyrae populations in Sgr and the LMC. This similarity is based on comparisons of the 
period distributions, the period-amplitude diagram, and, to a lesser extent, the period distributions of RRd stars. \\
 \item We estimate the total number of RRab stars to be $\sim$4\,200 in the main body of Sgr. If Sgr has lost 50$\%$ of its mass since its 
formation, the total number of RRab stars associated with Sgr would be $\sim$8\,400. If RRab stars trace light, the above estimates would 
correspond to an integrated V-magnitude of $-$13.9mag for the main body, and $-$14.7mag for the whole system.
\end{enumerate}
\indent The stellar population of Sgr has a high metallicity dispersion. This was first suggested from interpretation of the 
Color-Magnitude 
diagram of Sgr (Marconi et al. \cite{marconi}, Bellazzini et al. \cite{bel1}) and later confirmed by spectroscopic observations 
with values 
ranging from $-$1.4dex up to solar abundance (Smecker-Hane \& McWilliam \cite{shw}, Bonifacio \cite{bonifacio}). It seems that the 
RR Lyrae population also presents a high dispersion of $\sim$1dex around [Fe/H]$\simeq -$1.6, suggesting that the early star-formation 
process in Sgr was complex. Sgr could thus have formed on a large scale with star bursts occuring in several gaseous clumps of different 
metallicities. Another scenario would be a continuous star formation with progressive metal enrichment.  \\
\indent The similarity of RR Lyrae populations between Sgr and the LMC is indicative of 
 similar horizontal branch morphologies. This similarity implies that the parameter(s) driving the HB morphology are similar in both 
systems. Furthermore, since RR Lyraes represent the old metal-weak population of these systems, this result 
suggests that the LMC and Sgr formed at the same epoch and in a similar environment with respect to the metal abundance. It is thus tempting to   
speculate that Sgr and the LMC had a common progenitor. In this picture, Sgr could correspond to a piece of the LMC pulled out during a 
collision with the Galaxy. Such a scenario could help to explain how a galaxy with an old stellar population can be observed on such 
a low orbit without being completely disrupted through Galactic tides. 
%
%
%
%
\begin{acknowledgements}
I am pleased to thank Christophe Alard for constant support and many useful suggestions during the elaboration of this paper.
\end{acknowledgements}

\end{document}